\documentstyle[12pt,epsfig]{article}

\begin{document} 

\title{The Mesonic Fluctuations and Corrections in the Chiral Symmetry Breaking Vacuum} 

\author{Mei Huang$^{1,2}$, Pengfei Zhuang$^{1}$, Weiqin Chao$^{2,3}$ \\ 
 {\small  $^1$ Physics Department, Tsinghua University, Beijing 100084, China}\\
 {\small  $^2$ CCAST, Beijing 100080, China}\\ 
   {\small $^3$ IHEP, Chinese Academy of Sciences, Beijing 100039, China   }}

\maketitle

\date{}

\maketitle

\begin{abstract}

The mesonic quantum fluctuations and their corrections on the chiral condensate 
and pion polarization function are investigated in the self-consistent 
scheme of SU(2) Nambu-Jona-Lasinio model by exactly calculating the 
next-to-leading order (NLO) Feynman diagrams in $1/N_c$ expansion.  
While the fluctuations and corrections depend strongly on 
the  meson's three-momentum cut-off  $\Lambda_M$,
no chiral symmetry
restoration is found since the cancellation between the NLO and the 
one-quark-loop diagram with quark mass deviated from the mean-field value. 
 
\end{abstract}

\newpage

\section{Introduction}

It is generally accepted that the Nambu-Jona-Lasinio (NJL) \cite{NJL} model offers 
a simple scheme to study the mechanism of spontaneous chiral symmetry breaking.
Most calculations in this model have been based on the mean-field approximation, 
i.e., quark self-energy in the leading order of $1/N_c$ expansion \cite{NJL123}.

Recently, more interests are paid on the NJL model beyond mean-field approximation.
The mesonic fluctuations are variously investigated 
in the bosonization formalism \cite{NPA}-\cite{babaev}  and 
in the Feynman diagram formalism of $1/N_c$  expansion \cite{zhuang}-\cite{birse}.  
\cite{NPA} and \cite{ripka} in bosonization formalism using meson-loop expansion
or saddle-point expanstion are equivalent to the self-consistent Feynman diagram 
formalism \cite{Ann} in $1/N_c$ expansion, and the chiral symmetry of the NJL
model can be preserved. 
While, \cite{kleinert} and \cite{babaev} generalize the NJL model to the nonliear-sigma 
approach for description of chiral fluctuations, and the authors in \cite{kleinert}
claimed that at $N_c=3$ the NJL model does not display spontaneous symmetry breaking 
dut to chiral fluctuations.

There are two different $1/N_c$ expansion schemes which can preserve chiral symmetry 
of the NJL model. 
One is the so-called "strict" $1/N_c$ expansion scheme through iterating 
the quark self-energy at Hartree approximation \cite{oertel} \cite{birse}.
The other is the self-consistent scheme with meson-loop expansion or saddle-point expansion, 
in which the next-to-leading order (NLO) contributions represent the mesonic quantum
fluctuations around the classical stationary chiral symmetry breaking vacuum 
\cite{NPA}\cite{ripka}\cite{Ann}. In both schemes,
the meson corrections to the mean-field approximation are included not only in 
the meson-loop contributions, but also in the one-quark-loop contributions 
through the corrected quark propagator. 

In the above mentioned chiral symmetric schemes, emphasis is put on discussing the
NLO or pure meson-loop contributions.  However, to investigate
whether chiral symmetry is restored in the vacuum, one should investigate the meson 
corrections to the mean-field approximation.  In this paper we investigate not only  
the mesonic fluctuations around the classical vacuum, but also the
mesonic corrections to the mean-field approximation 
in the chiral symmetric self-consistent scheme based on \cite{Ann},
we will exactly calculate the meson polarization function and the quark self-energy 
to the next-to-leading order in $1/N_c$ expansion.
We will show that, by using the pole approximation for internal meson propagators, 
the $1/N_c$ expansion in the self-consistent scheme is also "strict".
The momentum cut-off for quarks and mesons, i.e., $\Lambda_f$ and $\Lambda_M$,
have been introduced in non-covariant three-momentum regularization.  
The model parameters at different
values of $\Lambda_M$ have been determined by fitting the pion's mass $m_{\pi}=139 MeV$ 
and the pion's decay constant $f_{\pi}=93.2 MeV$ with an appropriate value of the current quark 
mass. The way to determine the parameters to the next-to-leading order here is different from that 
used in \cite{oertel} and \cite{birse}, where the parameters are fitted at leading order \cite{birse}
or partialy at leading order \cite{oertel}.

\section{Mean-field approximation}

The two-flavor NJL model is defined through the Lagrangian density,
\begin{eqnarray}
\label{lagr}
{\cal L} = \bar{\psi}(i\gamma^{\mu}\partial_{\mu}-m_0)\psi + 
  G[(\bar{\psi}\psi)^2 + (\bar{\psi}i\gamma_5{\bf {\vec \tau}}\psi)^2 ],
\end{eqnarray}
here $G$ is the effective coupling constant of dimension ${\rm GeV}^{-2}$,
$m_0$ is the current quark mass, assuming isospin degeneracy of  
$u$ and $d$ quarks, and $\psi, \bar{\psi}$ are quark fields with flavor, colour 
and spinor indices suppressed. 

The quark mass $m_q^0$ in mean-field approximation,  i.e.,  in Hartree approximation,
can be represented by
\begin{eqnarray}
\label{gap0}
m_q^0 = m_0 + \Sigma_H^{0}, 
\end{eqnarray}
where  $m_0$ is the current quark mass and $\Sigma_H^{0}$ is the quark self-energy of  
one-quark-loop contributions 
in mean-field approximation, which can be expressed as
\begin{eqnarray}
\label{sigmah0}
\Sigma_H^{0}=8iG^0N_cN_f m_q^0 \int \frac{d^4p}{(2\pi)^4}\frac{1}{p^2-(m_q^0)^2}.
\end{eqnarray}
For quantities in mean-field approximation, we indicate
them with a superscript "$0$". 

The quark condensate calculated in mean-field approxiamtion has a simple relation
with the quark self-energy 
\begin{eqnarray}
-<{\bar q} q>^0=\frac{\Sigma_H^0}{4G^0}.
\end{eqnarray}

Correspondingly,  the meson polarization functions in random-phase approximation
are 
\begin{eqnarray}
 \Pi_{M}^{0}(k)& = &4iN_cN_f \int \frac{d^4p}{(2\pi)^4}
\frac{1}{p^2-(m_q^0)^2} - 2iN_cN_f(k^2-4 \epsilon_M (m_q^0)^2) \nonumber \\
& & \int \frac{d^4p}{(2\pi)^4}\frac{1}{(p^2-(m_q^0)^2)((p+k)^2-(m_q^0)^2)} ,
\end{eqnarray}
with $M$ refering to $\pi$ or $\sigma$, and $\epsilon_{\pi}=0$ and $\epsilon_{\sigma}=1$.

The NJL parameters in mean-field approximation can be fixed as $\Lambda_f= 637.7 MeV$, 
and $G^0 \Lambda_f^2 =2.16 $,  by choosing  an appropriate current quark mass $m_0=5.5 MeV$ 
and through  fitting the pion's mass $m_{\pi}=139MeV$ and decay constant $f_{\pi}=93.2 MeV$.
The corresponding quark mass is $m_q^0=330 MeV$ and  chiral 
condensate $(-<{\bar q} q>^0)^{1/3}=248 MeV$.

\section{Mesonic fluctuations and  corrections on chiral condensate}
~
The chiral symmetric self-consistent scheme  is described in details in \cite{NPA} \cite{ripka}
in effective action formalism and in \cite{Ann} by using Feynman diagramatic method. 

Including current quark mass $m_0$,  the gap equation for quark mass can be expressed as
\begin{eqnarray}
\label{gap}
m_q = m_0 + \Sigma_H + \Sigma^{fl}, 
\end{eqnarray}
where $\Sigma_H$ and $ \Sigma^{fl}$ are the leading order (LO) and next-to-leading order (NLO)
 of quark 
self-enery in $1/N_c$ expansion,  which can be read directly from the Feynman diagrams
in Fig.1.

The expression of $\Sigma_H$ is the same as that of $\Sigma_H^{0}$,  except  that the quark 
mass $m_q^0$ in Eq. (\ref{sigmah0})  should be replaced by $m_q$, which is the new saddle-point
of the effective action including mesonic fluctuations \cite{NPA} \cite{ripka}.
  
The NLO quark self-energy can be written directly from the
diagrams in Fig.1 
\begin{eqnarray}
\label{self1}
& & \Sigma^{fl}=-8GN_cN_f m_q \int\frac{d^4pd^4q}{(2\pi)^8}
  [ (-i D_{\pi}(q))
   (\frac{3}{(p^2-m_q^2)^2} \nonumber \\
& &   -\frac{3q^2}{(p^2-m_q^2)^2((p+q)^2-m_q^2)}  )
   +(-i D_{\sigma}(q))
   (\frac{1}{(p^2-m_q^2)^2}  \nonumber \\
& &  + \frac{2}{(p^2-m_q^2)((p+q)^2-m_q^2)}
   -\frac{q^2-4m_q^2}{(p^2-m_q^2)^2((p+q)^2-m_q^2)})] ,
\end{eqnarray} 
where  $-i D_{M}(q)$ indicate the internal meson propagator, which is in
random-phase-approximation, i.e., strings of single quark loops. 
At this position, it is necessary to note that in \cite{ripka}, the NLO quark self-energy
is seperated into two parts: the first term of meson propagator in RPA
and the remaining ring diagrams.

Evaluating quark loop and meson-loop integrals, one can get a simple relation between the 
quark condensate and  the constituent quark mass $m_q$,
\begin{eqnarray}
-<{\bar q} q>=-(<{\bar q} q>_H+<{\bar q} q>^{fl})=
                  \frac{\Sigma_H+\Sigma^{fl}}{4G}.
\end{eqnarray}

The magnitude of the mesonic quantum fluctuations around the classical vacuum 
is defined as the ratio of the pure NLO contributions over the LO contributions
\begin{eqnarray}
R_{q}^{fl}= \Sigma^{fl} /  \Sigma_H,
\end{eqnarray}
which had been discussed carefully in \cite{ripka}.
  
And the  mesonic corrections on the chiral condensate is defined as the difference
between the value in and beyond mean-field approximation, i.e.
\begin{eqnarray}
\delta <{\bar q} q>= (-<{\bar q} q>)-(-<{\bar q} q>^0)=
\frac{\Sigma_H+\Sigma^{fl}}{4G}-\frac{\Sigma_H^0}{4G^0},
\end{eqnarray}
and the ratio of mesonic corrections is
\begin{eqnarray}
R_{q}^{cr} = \delta <{\bar q} q> / (-<{\bar q} q>^0).
\end{eqnarray}

\section{Mesonic fluctuations and corrections on pion polarization function}
~
Because the sigma meson is heavy,  it does not play an important role in vacuum, 
and the NLO contributions would not affect the properties of sigma meson.
Here we only discuss the mesonic corrections on pion polarization function, and 
just simply choose $m_{\sigma} \simeq 2 m_q$.

The pion polarization function  $\Pi_{\pi}(k)$  to next-to-leading order  is shown in Fig. 1,  
and can be written as 
\begin{eqnarray}
\Pi_{\pi}(k)=\Pi_{\pi}^{(RPA)}(k)+ \Pi_{\pi}^{fl}(k),
\end{eqnarray}
where $\Pi_{\pi}^{(RPA)}$ and $ \Pi_{\pi}^{fl}$ are leading and subleading
contributions. 
The expression for $\Pi_{\pi}^{(RPA)}$ is the same as $\Pi_{\pi}^0$ except
that the quark mass $m_q^0$ replaced by $m_q$. 
And the contributions of sub-leading order to pion polarization function include three 
diagrams,   
\begin{eqnarray}
 \Pi_{\pi}^{fl} =\delta\Pi_{\pi}^{(b)} + \delta\Pi_{\pi}^{(c)} 
                    + \delta\Pi_{\pi}^{(d)} .
\end{eqnarray}
Each diagram can be evaluated as following:  
\begin{eqnarray}
\label{pib}
& & \delta\Pi_{\pi}^{(b)}(k) = 2N_cN_f\sum_{M=\pi, \sigma}\int\frac{d^4qd^4p}{(2\pi)^8}
(-i D_{M}(q))\nonumber \\
& & [\frac{1}{(p^2-m_q^2)((p+q-k)^2-m_q^2)}
+\frac{1}{((p+q)^2-m_q^2)((p-k)^2-m_q^2)} \nonumber \\
& &-\frac{k^2(q^2-\epsilon_M 4 m_q^2)}{(p^2-m_q^2)((p+q)^2-m_q^2)
((p-k)^2-m_q^2)((p+q-k)^2-m_q^2)}] ,
\end{eqnarray}
\begin{eqnarray}
& &  \delta\Pi_{\pi}^{(c)}(k) = -4N_cN_f\sum_{M=\pi,\sigma}
\int\frac{d^4qd^4p}{(2\pi)^8}\lambda_M(-i D_{M}(q)) 
\nonumber \\
& &[\frac{1}{((p+q)^2-m_q^2)((p-k)^2-m_q^2)} + 
\frac{k^2(q^2-\epsilon_M 4 m_q^2)}{(p^2-m_q^2)^2((p+q)^2-m_q^2)((p-k)^2-m_q^2)} 
 \nonumber \\
& & +\frac{1}{(p^2-m_q^2)^2} + 
 \frac{2k\cdot q}{(p^2-m_q^2)((p+q)^2-m_q^2)((p-k)^2-m_q^2)} \nonumber \\
& &-\frac{k^2}{(p^2-m_q^2)^2((p-k)^2-m_q^2)} 
- \frac{(q^2-\epsilon_M 4 m_q^2)}{(p^2-m_q^2)^2((p+q)^2-m_q^2)}],
\end{eqnarray}
with the degeneracy $\lambda_\pi=3$, $\lambda_{\sigma}=1$ and
\begin{eqnarray}
& & \delta\Pi_{\pi}^{(d)}(k)=i\int\frac{d^4q}{(2\pi)^4}
(-i D_{\pi}(q))(-iD_{\sigma}(q-k)) \nonumber \\
& & [\int\frac{d^4p}{(2\pi)^4}\frac{8m_qN_cN_f(k \cdot q-(p^2-m_q^2))}
{(p^2-m_q^2)((p+q)^2-m_q^2)((p+k)^2-m_q^2)}]^2.
\end{eqnarray}

The magnitude of mesonic fluctuations on pion polarization function is defined as 
the ratio of the next-to-leading order contribution over the leading order ones
\begin{eqnarray}
R_{\pi}^{fl}= \Pi_{\pi}^{fl} / \Pi_{\pi}^{(RPA)}.
\end{eqnarray}
Considering mesonic fluctuations the corrections on pion polarization function is 
\begin{eqnarray}
\delta \Pi_{\pi}=\Pi_{\pi}-\Pi_{\pi}^{0}, 
\end{eqnarray}
and the magnitude of mesonic corrections on pion polarization can be defined as
\begin{eqnarray}
R_{\pi}^{cr}=\delta \Pi_{\pi}/  \Pi_{\pi}^{0}.
\end{eqnarray}

At last,  we should discuss the formalism of the internal meson propagators.
It has been discussed in several papers that the $1/N_c$ expansion in the self-consistent scheme 
is not strict \cite{huang} \cite{oertel}. It has been pointed out that  the exact next-to-leading 
order quark self-enargy 
includes higher order contributions because the RPA  internal meson propagators include
any higher order $O(1/N_c^n)$  contributions.
In our calculations, we choose the  internal meson propagators
\begin{eqnarray}
-iD_{M}(q)=\frac{2iG}{1-2G\Pi_{M}^{(RPA)}(q)}
\end{eqnarray}
in pole aprroximation 
\begin{eqnarray}
-iD_M(q)=-i \frac{[g_{M qq}^{(RPA)}(q)]^2}{q^2-m_M^2},
\end{eqnarray}
 with the pion-quark-antiquark coupling constant
\begin{eqnarray}
\label{couple}
g_{M qq}^{(RPA)}(q) = (\partial \Pi_{M} ^{(RPA)}(q)/\partial q^2)^{-2}.
\end{eqnarray}
The internal meson propagators in pole approximation will be always in the 
order of $O(1/N_c)$,  which ensure the $1/N_c$ expansion is "strict" in the 
self-consistent scheme.

\section{Numerical Results }
~
Our numerical results are based on solving three equations,
the gap equation for quark Eq.(\ref{gap}),  the equation for pion mass  
\begin{eqnarray}
\label{pole}
1-2G\Pi_{\pi}(k^2=m_{\pi}^2)=0,
\end{eqnarray}
where the pion polarization function $\Pi_{\pi}$ is calculated completely 
without any approximation, and the equation for pion decay constant 
\begin{eqnarray}
\label{onshell}
\frac{m_{\pi}^2f_{\pi}}{g_{\pi qq}} = \frac{m_0}{2G},
\end{eqnarray}
where the total coupling constant $g_{\pi qq}$  is given by the residue of the total
pion polarization at the pole
\begin{eqnarray}
\label{couple}
g_{\pi qq}^{-2} = \partial \Pi_{\pi} / \partial k^2|_{k^2=m_{\pi}^2}.
\end{eqnarray}

To evaluate the integrals, we  introduce two three-momentum cut-offs $\Lambda_f$ and
$\Lambda_M$ for quarks and mesons in the non-covariant regularization, in order that the fixed
model parameters can also be used at finite temperature and density in the future studies.
The three model parameters $m_0$, $G$ and $\Lambda_f$ are determined varying with 
$\Lambda_M$ by fitting
the pion mass $m_{\pi}=139 MeV$ and 
pion decay constant $f_{\pi}=93.2MeV$ and taking an appropriate current quark mass 
$m_0=5.5 MeV$. 

Different from \cite{oertel} and \cite{birse}, we did not use 
the leading order of the Gell-Mann-Oakes-Renner (GOR) relations 
\begin{equation}
-m_0<{\bar q} q>=m_{\pi}^2f_{\pi}^2+O(m_{\pi}^4)+ \dots
\end{equation}
as a constraint,
because we did not use  external moentum expansion for pions in our calculations.
In addition, we did not  use chiral expansion for pion mass and pion decay constant. 
As for the way to determine the model parameters in the calculations
to the next-to-leading order, we don't think it is self-consistent to fit the parameters at
\cite{birse} or partially at \cite{oertel} leading order.

Our numerical results are shown in Fig.2 and Fig.3 as a function of $\Lambda_M /m_q$ . 
It is noticed that
our results always start at about $\Lambda_M/m_q=1$.  This is because for any value
of $\Lambda_M$, the solutions of $m_q$ satisfies the relation $\Lambda_M/m_q >1$. 

In Fig. 2.a, the total (solid circles) and LO (stars) quark condensate,
$-<{\bar q}q>^{1/3}$ and $-<{\bar q}q>_H^{1/3}$, 
are plotted as a function of $\Lambda_M/m_q$,  
the value at $\Lambda_M/m_q=0$ corresponds 
to the mean-field aprroximation $-(<{\bar q}q>^0)^{1/3}$. 
In Fig.2.b, the magnitude of mesonic fluctuations $R_q^{fl}$ (stars)
and mesonic corrections $R_q^{cr}$ (solid circles) to quark condensate
are plotted as a function of  $\Lambda_M/m_q$. 

In Fig. 3.a, the total (solid circles) and LO (stars) 
pion polarization function, $\Pi_{\pi}(m_{\pi}^2)$ and 
$\Pi^{(RPA)}_{\pi}(m_{\pi}^2)$, are
plotted as a function of $\Lambda_M/m_q$,  
and the value at $\Lambda_M/m_q=0$ is the pion polarization function 
in mean-field approximation $\Pi^0_{\pi}(m_{\pi}^2)$. 
And the magnitude of mesonic fluctuations $R_{\pi}^{fl}$ (stars)
and mesonic corrections $R_{\pi}^{cr}$ (solid circles) on pion 
polarization function are plotted as a function of  $\Lambda_M/m_q$ in Fig.3.b.

It is found that in the region $1<\Lambda_M/m_q <2.3$ with 
$270 MeV < m_q < 440 MeV$ and $270 MeV<\Lambda_M<1000MeV$,
both the LO and total quark condensate and pion polarization function 
are smaller than their mean-field values, which is refelcted in the negative meson 
corrections. Here the meson corrections reducing the mean-field quark condensate
agree with the results in Fig.7 of \cite{NPA}. 

The mesonic fluctuations, i.e., the pure NLO contributions
of the chiral condensate and pion polarization function
can be characterized by the meson-momentum 
cut-off $\Lambda_M$, larger $\Lambda_M$ results in larger meson-loop contibutions. 
The NLO contributions are very small till $\Lambda_M/m_q=1.5$ 
with $m_q=m_q^0=330 MeV$,  
then increase fast. When $\Lambda_M/m_q>2.3$, corresponding to
$\Lambda_M >1000 MeV$, the mesonic fluctuations are larger than
$40 \%$, which means that the $1/N_c$ expansion around the 
stationary classical vacuum could not be used any more in this case.
This is reasonable for only scalar and pseudo-scalar mesons involved.
It is noticed that the NLO contributions enhancing the LO quark 
condensate is the same as that in \cite{NPA} and \cite{ripka}. 

However, the negative mesonic corrections of the chiral condensate and 
pion polarization function do not decrease monotonously with the increasing 
$\Lambda_M/m_q$, they reach the minimum $-35 \%$ and $-40 \%$ at 
about $\Lambda_M/m_q=2$ with $m_q=370 MeV$, then
increase fast. The reason is that the meson corrections to the mean-field approximation 
include two parts:  the pure meson-loop or NLO contributions
and the deviation of the one-quark-loop from the mean-field calculation with
the quark mass $m_q^0$. When $1<\Lambda_M/m_q<1.5$, corresponding 
to $m_q <m_q^0$,
the pure meson-loop contributions are negative but very small, and the meson corrections 
come only from the change of quark mass in the one-quark-loop part. 
When $\Lambda_M/m_q>1.5$, corresponding to $m_q>m_q^0$,
the pure meson-loop contributions are contrary to the contributions from the deviation of the
one-quark-loop contributions. At about $\Lambda_M/m_q=2.3$, the two parts cancel to each other
completely, and the meson corrections vanish. 

It is the suppresion from the LO contributions and the enhancement from  
the NLO contributions, that the quark condensate in Fig.2 $a$ and the pion polarization function 
in Fig.3 $a$ do not decrease continuously with increasing $\Lambda_M/m_q$.
In \cite{kleinert}, the authors claim that chiral symmetry will restore in vacuum
by quantum fluctuations. One reason to get this strange conclusion is the lack of meson-loop
contribution in the calculations in \cite{kleinert}. This is also pointed out in \cite{ripka}.

It is necessary to point out that the cancellation between the NLO contributions and
the deviation of one-quark-loop contributions from the mean-field approximation 
is not the same as the cancellation between the exchange term and the remaining ring 
diagrams in \cite{ripka}. In the latter case, the meson-loop contributions to quark self-energy
are seperated into two parts, the exchange diagram and the ring diagram. It is found that the
two contributions are contrary, and the exchange term is dominant, which is very important 
to ensure that the NLO contribution is positive to the LO values. 
Our cancellation is also not the same as the cancellation between the two NLO diagrams in \cite{birse},
where the cancellation induces the NLO contribtions to LO are very small.
The cancellation inside the NLO contributions in \cite{ripka} and \cite{birse}
can not reflect the convergence of the total mesonic corrections to the mean-field 
approximation, because the mesonic corrections is also included in the one-quark-loop 
diagram due to the change of quark mass.

\section{Conclusions}

In conclusion,  the magnitude of mesonic fluctuations and their corresponding corrections
on chiral condensate and pion polarization function are investigated in a self-consistent
scheme by exactly calculating the Feynman diagrams to next-to-leading order
of $1/N_c$ expansion. 

It is found that the magnitude of the mesonic fluctuations around the classical vacuum, i.e., 
the pure meson-loop contributions, can be characterized by the meson-momentum 
cut-off $\Lambda_M$, the larger $\Lambda_M$, the larger mesonic fluctuations.
The NLO contributions enhance the LO values at the stationary
saddle-point. However, the LO contributions reduce the values in the mean-field approximation.
The two contributions to the mean-field approximation  are contrary and 
cancel to each other. 

When $\Lambda_M>1000 MeV$, the mesonic fluctuations will be larger than 
$40 \%$, which means that the meson-loop expansion or $1/N_c$ expansion could not be 
used any more.  In the case of only scalar and pseudo-scalar mesons involved,
$\Lambda_M< 1000MeV$ is reasonable,  otherwise,  heavier vector mesons 
should be considered.

Even though we can not answer the question 
how much the magnitude of the mesonic fluctuations and
corrections are, because the model parameters are not uniquely determined, 
we can conclude that the cancellation between the NLO contributions and
the one-quark-loop with quark mass deviation from 
the mean-field value induces an appropriate mesonic corrections to the 
mean-field approximation, and no chiral symmetry restoration 
or instabilities of the vacuum is found. 
 
\section*{Acknowledgements}
~
This work was supported in part by the NSFC under Grant No. 19925519 and
by the Major State Basic Research Development Program under 
Contract No. G2000077407.

\begin{figure}[ht]
\vspace*{-2truecm}
\centerline{\epsfxsize=16cm\epsffile{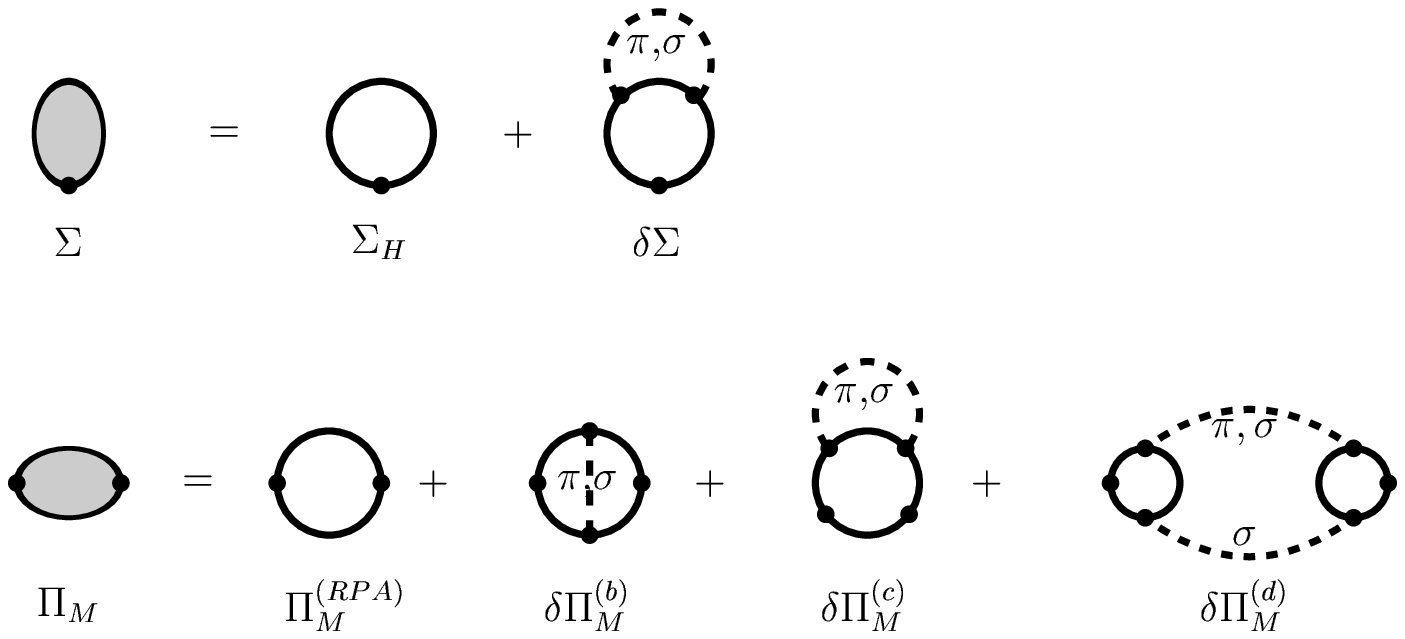}}
\vspace*{-10truecm}
\caption
{ Quark self-energy $\Sigma$ and pion polarization
function $\Pi_M$ in the quark and pion propagators. 
$\Sigma_H$ and $ \Sigma^{fl}$ are the leading and subleading contributions to 
the quark mass. $\Pi^{(RPA)}_{\pi}$ and $\delta \Pi^{(b,c,d)}_{\pi}$ are 
the leading 
and subleading order contributions to pion polarization function. 
The heavy solid lines
indicate the constituent quark propagator,  
and the heavy dashed lines represent 
$\pi$ or $\sigma$ propagator $-{\rm i} D_{M}(q)$ in RPA
approximation. }
\label{kernel_fig}
\end{figure}
\newpage

\begin{figure}[ht]
\vspace*{-4truecm}
\centerline{\epsfxsize=13.5cm\epsffile{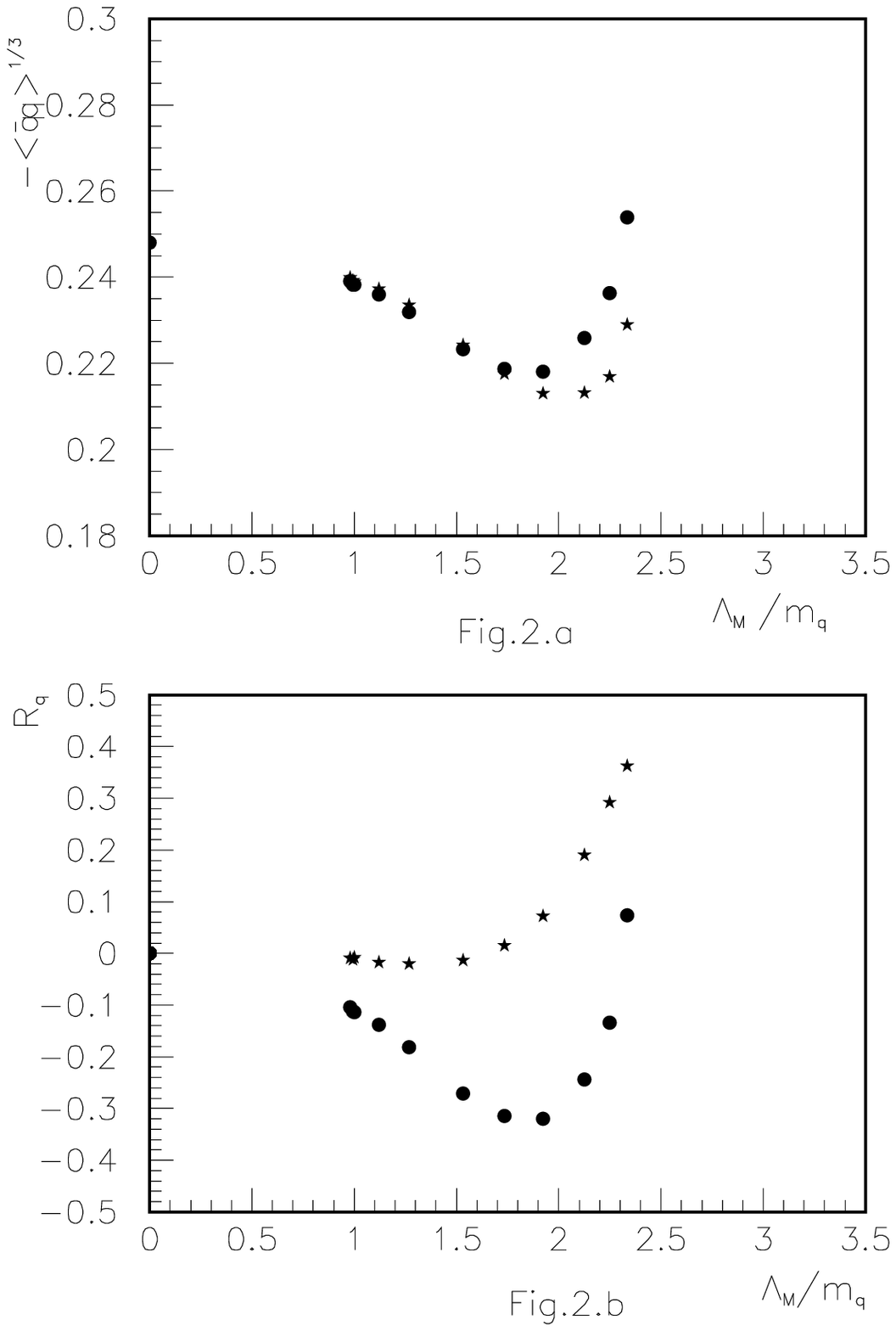}}
 
\caption
{Quark condensate (a) and the magnitude of mesonic fluctuations and 
correction on  chiral condensate (b) as 
a function $\Lambda_M/m_q$ .
The solid circles and stars in (a) correspond to $-<{\bar q}q>^{1/3}$
and $-<{\bar q}q>_H^{1/3}$ respectively, and in (b) correspond to
the mesonic corrections $R_q^{cr}$ and meson fluctuations $R_q^{fl}$ 
on quark condensate respectively.}

\end{figure}
\newpage

\begin{figure}[ht]
\vspace*{-4truecm}
\centerline{\epsfxsize=13.5cm\epsffile{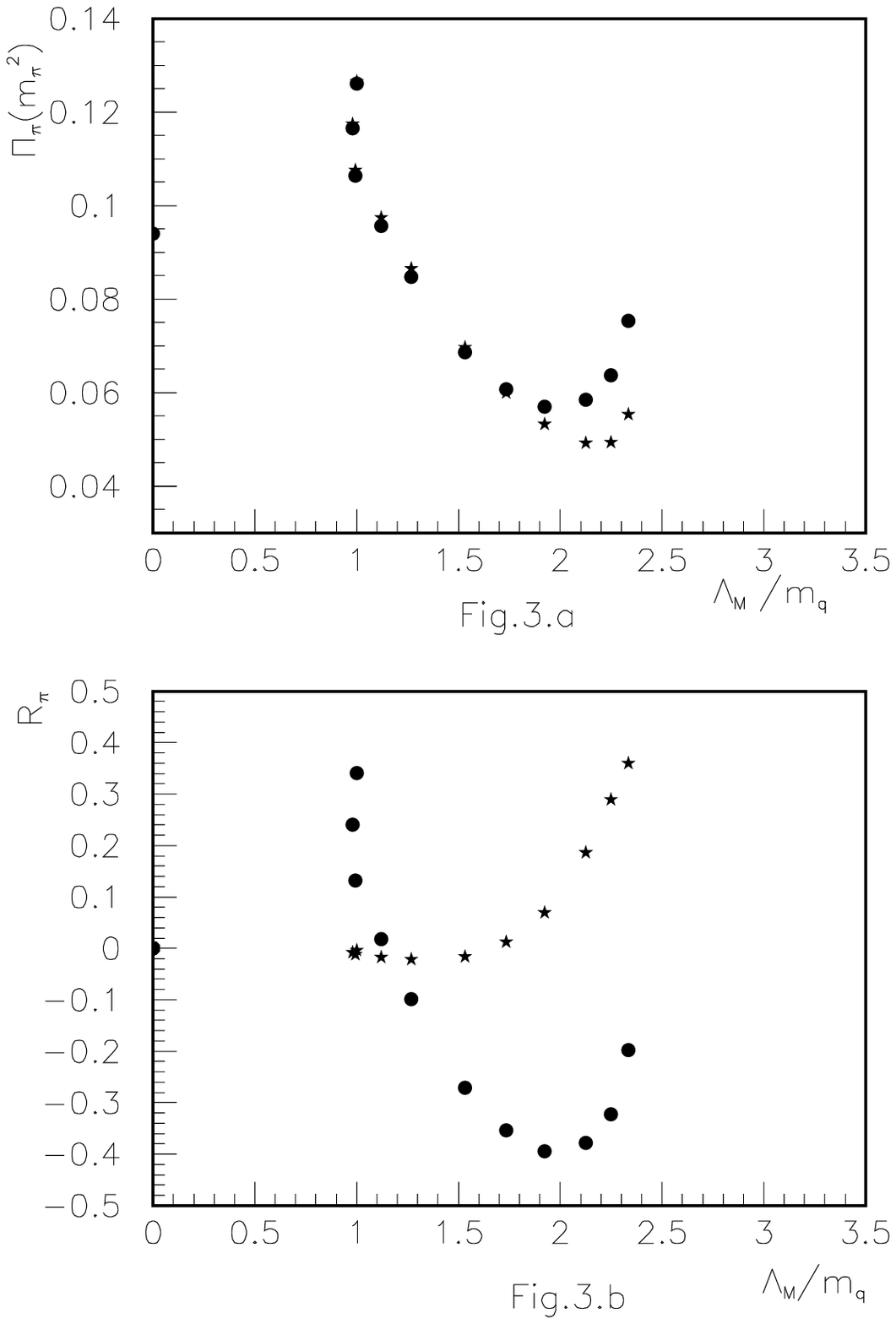}}
 
\caption
{Pion polarization function (a) and the magnitude of mesonic fluctuations and 
correction on pion polarization function (b) as 
a function of $\Lambda_M/m_q$ .
The solid circles and stars in (a) correspond to $\Pi_{\pi}(m_{\pi}^2)$ 
and $\Pi_{\pi}^{(RPA)}(m_{\pi}^2)$  respectively, and in (b) correspond to
the mesonic corrections $R_{\pi}^{cr}$and meson fluctuations $R_{\pi}^{fl}$ 
on pion polarization function respectively.}

\end{figure}

\end{document}